\documentclass[twocolumn,secnumarabic,amssymb, nobibnotes, aps, prd, superscriptaddress]{revtex4}

\usepackage[dvips]{graphicx}

\begin{document}

\title{Charge transfer in H + He$^+$ and H$^+$ + He collisions in excited states}

\author{J. Loreau}
\email{jloreau@ulb.ac.be}
\affiliation{Service de Chimie Quantique et Photophysique, Universit\'e libre de Bruxelles (ULB) CP 160/09, 1050 Brussels, Belgium}
\author{S. Ryabchenko}
\affiliation{Northern (Arctic) Federal University, 17 Severnaya Dvina Emb., 163002 Arkhangelsk, Russia}
\author{J. M. Mu\~noz Burgos}
\affiliation{Astro Fusion Spectre, LLC, San Diego, CA 92122, USA}
\author{N. Vaeck}
\affiliation{Service de Chimie Quantique et Photophysique, Universit\'e libre de Bruxelles (ULB) CP 160/09, 1050 Brussels, Belgium}


\begin{abstract}
We present a theoretical study of charge transfer in collisions of excited ($n=2,3$) hydrogen atoms with He$^+$ and in collisions of excited ($n=2,3$) helium atoms with H$^+$, extending the results of {\it Phys. Rev. A} {\bf 82}, 012708 (2010). A combination of quantum-mechanical and semi-classical approaches is employed to calculate the charge-exchange cross sections at collision energies from 0.1 eV/u up to 1 keV/u. These methods are based on accurate {\it ab initio} potential energy curves and non-adiabatic couplings for the molecular ion HeH$^+$. 

Charge transfer can occur either in singlet or in triplet states, and the differences between the singlet and triplet spin manifolds are discussed. The dependence of the cross section on the quantum numbers $n$ and $l$ of the initial state is demonstrated. The isotope effect on the charge transfer cross sections, arising at low collision energy when H is substituted by D or T, is investigated. 
Finally, the impact of the present calculations on models of laboratory plasmas is discussed. 

\end{abstract}

\maketitle

\section{Introduction}

Charge transfer processes occur in many environments, from laboratory and astrophysical plasmas to biomolecules. In astrophysics, charge transfer in collisions between solar wind ions with neutral atoms is responsible for X-ray and EUV emissions from comets and other solar system bodies \cite{Bhardwaj2007,Bodewits2005} as well as from supernova remnants \cite{Cumbee2014}, galaxy clusters \cite{Gu2015} or star-forming galaxies \cite{Liu2012a}.

Charge transfer in ion-atom collisions also has substantial applications in the study of laboratory plasmas, as powerful diagnostics tools of magnetic fusion plasmas are based on this process. 
These methods take advantage of the presence of impurities in the plasma due to erosion of the inner wall of the tokamak. 
An example is provided by Charge Exchange Spectroscopy \cite{Isler1994}, which uses an energetic neutral beam of hydrogen or deuterium that interacts with the plasma. The capture of electrons from the neutral atoms by highly-charged impurity ions and the subsequent radiative decay cascade can be employed to infer key physical properties of the plasma such as impurity temperatures and densities. A fast thermal neutral beam of helium can also be used \cite{Agostini2015}, allowing a greater penetration in the plasma due to the higher ionization energy compared to hydrogen. The comparison between experimental and modelled line intensities gives access to the electron density and temperature.
On the other hand, non-invasive diagnostics methods include the observation of emission lines from recombining impurities. Among these, helium-like ions emission lines in particular are used as an efficient tool to establish a diagnostics of the plasma \cite{Bertschinger2005,Rosmej2006,Munoz2012}. Helium itself, being produced by the fusion reaction, can be used to study the conditions of the plasma edge and divertor regions.
The limitations of the models based on both methods are the availability of accurate collisional data such as charge transfer cross sections for collisions of hydrogen or helium atoms with impurity ions in various ionization stages, and theoretical and experimental efforts have focused on producing such data for various ions.

In the present work we focus on the charge transfer processes
\begin{equation}
\label{CX_1}
\textrm{He}^{+} (1s) + \textrm{H}(nl) \rightarrow \textrm{He} (1sn^{\prime}l^{\prime}\  ^{1,3}L) + \textrm{H}^{+} 
\end{equation}

\begin{equation}\label{CX_2}
\textrm{He} (1snl \ ^{1,3}L) + \textrm{H}^+ \rightarrow \textrm{He}^+ (1s) + \textrm{H}(n^{\prime}l^{\prime}) 
\end{equation}  
where $n=2-3$ and $n^\prime=1-3$.

The processes (\ref{CX_1}) and (\ref{CX_2}) have been extensively studied both experimentally and theoretically for $n=1$, i.e. when the neutral atom is initially in the ground electronic state. Such collisions provide one of the simplest systems to understand the basic mechanisms of charge transfer, and the accurate determination of collision cross sections is required for plasma modeling. The results cover the range from 10 eV/u up to several MeV/u. Various methods have been employed to calculate the charge transfer cross section, including purely quantum-mechanical methods at low energy ($\lesssim$ 1 keV/u), semi-classical approaches at intermediate energy (100 eV/u $\lesssim E \lesssim$ 10 keV/u), and methods based on atomic wavefunctions at high energies ($E \gtrsim$ 10 keV/u). 

Charge transfer in He$(1s^2)$ + H$^+$ collisions has been the subject of many experimental \cite{Schwab1987,Shah1989,Gudmundsson2010,Jaecks1965,Williams1967,Hughes1971,Crandall1971,Rudd1983,Hippler1987,Cline1994,Schoffler2009,Kim2012} and theoretical \cite{Jackson1953,Bransden1966,Winter1974,Jain1987,Chan1979,Slim1990,Slim1991,Mancev2003,Mancev2010,Ghanbari2011,Belkic2008} studies, in which the relatives contributions of electron capture into the ground and excited electronic states was assessed.
Recently process (\ref{CX_2}) was re-examined and it was demonstrated that the total and state-to-state cross sections can be obtained over an energy range from 10 eV/u up to 10 MeV/u by combining different theoretical approaches, thereby providing a recommended cross section for this reaction \cite{Loreau2014a}.
Charge transfer in He$^+(1s)$ + H$(1s)$ collisions has also been the focus of numerous experimental and theoretical studies \cite{Olson1977b,Macias1983,Hvelplund1982,Jackson1992,Ermolaev1994,Kuang1995,Mancev2007,Samanta2012,Liu2017}.

However, charge transfer processes in which the H or He are initially in an excited electronic state can have major consequences on the outcome of the laboratory plasma models. While the population of these excited states is small, the charge transfer cross sections are orders of magnitude larger than for the ground state, which can strongly affect the He emission lines. 

In two previous papers \cite{Loreau2010b,Loreau2011c} we examined process (\ref{CX_1}). 
In the first \cite{Loreau2010b} we computed the charge transfer cross sections for singlet states in the energy range 0.25-150 eV/u. 
In the second paper \cite{Loreau2011c}, we focused on the isotope effect in the $n=2$ singlet and triplet states that occurs when T is substituted to H.
The aim of the present paper is to extend these results as follows: 
({\it i}) We compute the cross sections for reaction (\ref{CX_1}) both in singlet and triplet $n=2,3$ states and provide spin-averaged cross sections; 
({\it ii}) We extend our previous results over the energy range 0.1-1000 eV/u;
({\it iii}) We calculate the cross sections for process (\ref{CX_2}) over the same energy range;
({\it iv}) We investigate the isotope effect on processes (\ref{CX_1}) and (\ref{CX_2}) where H is substituted by deuterium or tritium for all $n=2,3$ states.

The process (\ref{CX_2}) has been previously studied theoretically for $n=2,3$ \cite{Chibisov2001,Chibisov2002} in the energy range between 2 eV/u and 5 keV/u using an atomic orbital semi-classical approach, but large discrepancies were observed between these semi-classical results and calculations performed with a quantum wave packet approach in the singlet manifold \cite{Loreau2010b}. The process (\ref{CX_2}) for He($1s2s\ ^{1,3}S)$ was also investigated by Liu {\it et al.} at collision energies in the range 2-200 keV/u by means of the two-centre atomic orbital close-coupling \cite{Liu2012b}, showing large discrepancies with Refs. \cite{Chibisov2001,Chibisov2002}.

This article is organized as follows. In Section II we recall the main features of our theoretical approach to study the charge transfer reaction. In Section III we discuss the results for processes (\ref{CX_1}) and (\ref{CX_2}) and the influence of the isotope effect. In Section IV we summarize our results and we discuss the applications on the calculations presented here to the modelling of fusion plasmas and their diagnostic.

\section{Theoretical methods}

Three methods were employed in order to calculate the charge transfer cross sections. 
At collision energies below 100 eV/u, we used a quantum-mechanical (QM) approach based on Gaussian wave packet propagation in the diabatic representation. This time-dependent method has been described in detail elsewhere \cite{Vaeck1999,Baloitcha2001,Loreau2010b}. At the lowest energies, where time-dependent methods face numerical issues due to long propagation times, we employed a time-independent approach. At high collision energies, the QM methods become intractable due to the large number of contributing partial waves. At collision energies above 20 eV/u, we calculated the charge transfer cross sections by means of an eikonal semi-classical (SC) method in the impact parameter approximation \cite{Allan1990}.
In a previous study of H$(1s)$ + He$^+(1s)$ collisions we showed that the cross sections obtained with the quantum-mechanical and the eikonal methods provide equivalent results in the energy range around 700 eV/u. In the present case, we observed that the overlap between the two methods occurs at much lower energies, around 50 eV/u. In the following, the cross sections that will be presented were obtained using the QM and SM methods for energies below and above 50 eV/u, respectively. All the cross sections presented in this work are available as supplementary material \cite{suppmat}

The three methods that were employed are all based on a molecular description of the collision. The {\it ab initio} potential energy curves (PECs) of the molecular ion HeH$^+$ corresponding to the charge transfer processes (\ref{CX_1}) and (\ref{CX_2}) were calculated and discussed previously \cite{Loreau2010a,Loreau2010c,Loreau2011b,Sodoga2009,Loreau2013c}. For $n=1$ there are 2 $^1\Sigma^+$ and 1 $^3\Sigma^+$ states, for $n=2$ there are 4 $^{1,3}\Sigma^+$ and 2 $^{1,3}\Pi^+$ states, while for $n=3$ there are 6 $^{1,3}\Sigma^+$, 4 $^{1,3}\Pi^+$, and 2 $^{1,3}\Delta$ states. The total number of molecular states in our calculations is therefore 12 $^1\Sigma^+$, 11 $^3\Sigma^+$, 6 $^{1,3}\Pi^+$, and 2 $^{1,3}\Delta$ states. While charge transfer can occur in singlet and triplet states, we neglect the coupling between the spin manifolds so that they can be treated separately. The molecular states and their dissociation products and energies are summarized in Table \ref{table_states}.
The non-adiabatic couplings, which control the dynamics of the charge transfer process at the avoided crossings between the PECs, have been presented and analysed in Refs \cite{Loreau2010a,Loreau2011b}.

\begin{table}[h!]
\begin{center}
\begin{tabular}{lccc c }
			& $m$	& Colliding partners  	& Molecular states 		& Energy (hartree)		\\ \hline
$n=1$		& 1	& H$^+$ + He($1s^2 \ ^1S$) 	& $^1\Sigma^+$					& -2.90338589	\\
			& 2	& H($1s$) + He$^+(1s)$ 		& $^1\Sigma^+$					& -2.49954925	\\
$n=2$		& 3	& H$^+$ + He($1s2s\ ^1S$) 	& $^1\Sigma^+$					& -2.14577013	\\
			& 4	& H($2p$) + He$^+(1s)$ 		& $^1\Sigma^+$, $^1\Pi$				& -2.12474895	\\
			& 5	& H($2s$) + He$^+(1s)$ 		& $^1\Sigma^+$					& -2.12474895	\\
			& 6	& H$^+$ + He($1s2p\ ^1P$) 	& $^1\Sigma^+$, $^1\Pi$				& -2.12363793	\\
$n=3$		& 7	& H$^+$ + He($1s3s\ ^1S$) 	& $^1\Sigma^+$					& -2.06107975	\\
			& 8	& H$^+$ + He($1s3d\ ^1D$) 	& $^1\Sigma^+$, $^1\Pi$, $^1\Delta$	& -2.05542927	\\
			& 9	& H($3p$) + He$^+(1s)$ 		& $^1\Sigma^+$, $^1\Pi$				& -2.05534163	\\
			& 10	& H($3s$) + He$^+(1s)$ 		& $^1\Sigma^+$, $^1\Pi$, $^1\Delta$	& -2.05534163	\\
			& 11	& H($3s$) + He$^+(1s)$ 		& $^1\Sigma^+$					& -2.05534163	\\
			& 12	& H$^+$ + He($1s3p\ ^1P$) 	& $^1\Sigma^+$, $^1\Pi$				& -2.05495360	\\\hline
$n=1$		& 1	& H($1s$) + He$^+(1s)$ 		& $^3\Sigma^+$					& -2.49954925	\\
$n=2$		& 2	& H$^+$ + He($1s2s\ ^3S$) 	& $^3\Sigma^+$					& -2.17502848	\\
			& 3	& H$^+$ + He($1s2p\ ^3P$)	& $^3\Sigma^+$, $^3\Pi$				& -2.13296955	\\
			& 4	& H($2p$) + He$^+(1s)$	 	& $^3\Sigma^+$, $^3\Pi$				& -2.12474895	\\
			& 5	& H($2s$) + He$^+(1s)$ 		& $^3\Sigma^+$					& -2.12474895	\\
$n=3$		& 6	& H$^+$ + He($1s3s\ ^3S$) 	& $^3\Sigma^+$					& -2.06849764	\\
			& 7	& H$^+$ + He($1s3p\ ^3P$) 	& $^3\Sigma^+$, $^3\Pi$				& -2.05789144	\\
			& 8	& H$^+$ + He($1s3d\ ^3D$) 	& $^3\Sigma^+$, $^3\Pi$, $^3\Delta$	& -2.05544485	\\
			& 9	& H($3p$) + He$^+(1s)$ 		& $^3\Sigma^+$, $^3\Pi$				& -2.05534163	\\
			& 10	& H($3s$) + He$^+(1s)$ 		& $^3\Sigma^+$, $^3\Pi$, $^3\Delta$	& -2.05534163	\\
			& 11	& H($3s$) + He$^+(1s)$ 		& $^3\Sigma^+$					& -2.05534163	\\
\hline \hline
\end{tabular}
\caption{Atomic states involved in the charge transfer processes (\ref{CX_1}) and (\ref{CX_2}) and corresponding molecular states of HeH$^+$ in the singlet and triplet manifolds, with their asymptotic energies. See Ref. \cite{Loreau2010a} for details.}
\label{table_states}
\end{center}
\end{table}

There are various difficulties in treating processes (\ref{CX_1}) and (\ref{CX_2}).
First, there are several initial and final states in the same energy range. For a given $n>1$ the electronic states are very close in energy (and are degenerate asymptotically in the case of H($nl$)), which leads to strong non-adiabatic interactions (i.e., avoided crossings) between the PECs that describe the collision and thus large charge transfer cross sections. With increasing principal quantum number $n$ the energy differences between the atomic levels describing the initial and final states of processes (\ref{CX_1}) and (\ref{CX_2}) decrease and the charge transfer reaction becomes quasi-resonant. Additionally, this means that collisional excitation will also occur with large cross sections. 

Secondly, the non-adiabatic interactions take place in two distinct settings. Charge transfer can occur through non-adiabatic interactions at short-range, where the PECs are strongly interacting and numerous avoided crossings between the molecular states are present \cite{Loreau2010a}.
For the system under consideration, it can also occur at long range. The PECs of the molecular states of HeH$^+$ have different asymptotic behaviours according to whether they dissociate into He($1snl$) + H$^+$ or He$^+(1s)$ + H$(nl)$. In the first case, the PECs have an $R^{-4}$ form due to the ion-induced dipole interaction, with the exact dependence being controlled by the polarizability $\alpha$ of the excited state of the helium atom. However, in the second case the dominant contribution to the asymptotic PECs depends on $R^{-2}$ due to the first-order Stark effect for the degenerate hydrogen electronic states. This results in long-range avoided crossings, leading to charge transfer. 
This issue is however simplified by the fact that the long-range avoided crossings can be considered diabatic, {\it i.e.} we can assume that all the population is transferred at these crossing points. 
It should be noted, however, that for a given $n$ there is no Stark effect for the molecular state dissociating into H($nl$) + He$^+$ when $\vert\Lambda\vert=n-1$. In this particular case the PEC displays the same asymptotic $R^{-4}$ behavior as the He($1snl)$ + H$^+$ states. It follows that the PECs of the two $\Pi$ states (for $n=2$) and the two $\Delta$ states (for $n=3$) are almost parallel with a wide, Demkov-type non-adiabatic coupling \cite{Bransden1992}, leading to a different charge transfer dynamics.

\section{Results and discussion}

\subsection{Charge transfer in H($nl)$ + He$^+(1s)$ collisions}

\begin{figure}
\includegraphics[,width=.5\textwidth]{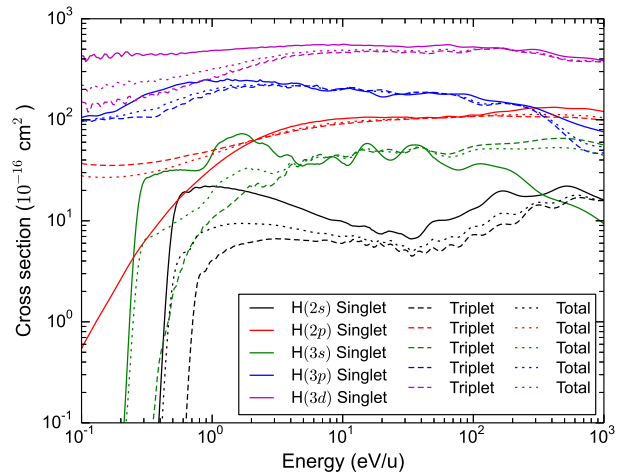}
\caption{Charge transfer cross sections for H($nl$) + He$^+ \rightarrow$ H$^+$ + He collisions in singlet and triplet states. The total (spin-averaged) cross section is also shown.}
\label{fig_comp_H_1}
\end{figure}

\begin{figure}
\includegraphics[,width=.5\textwidth]{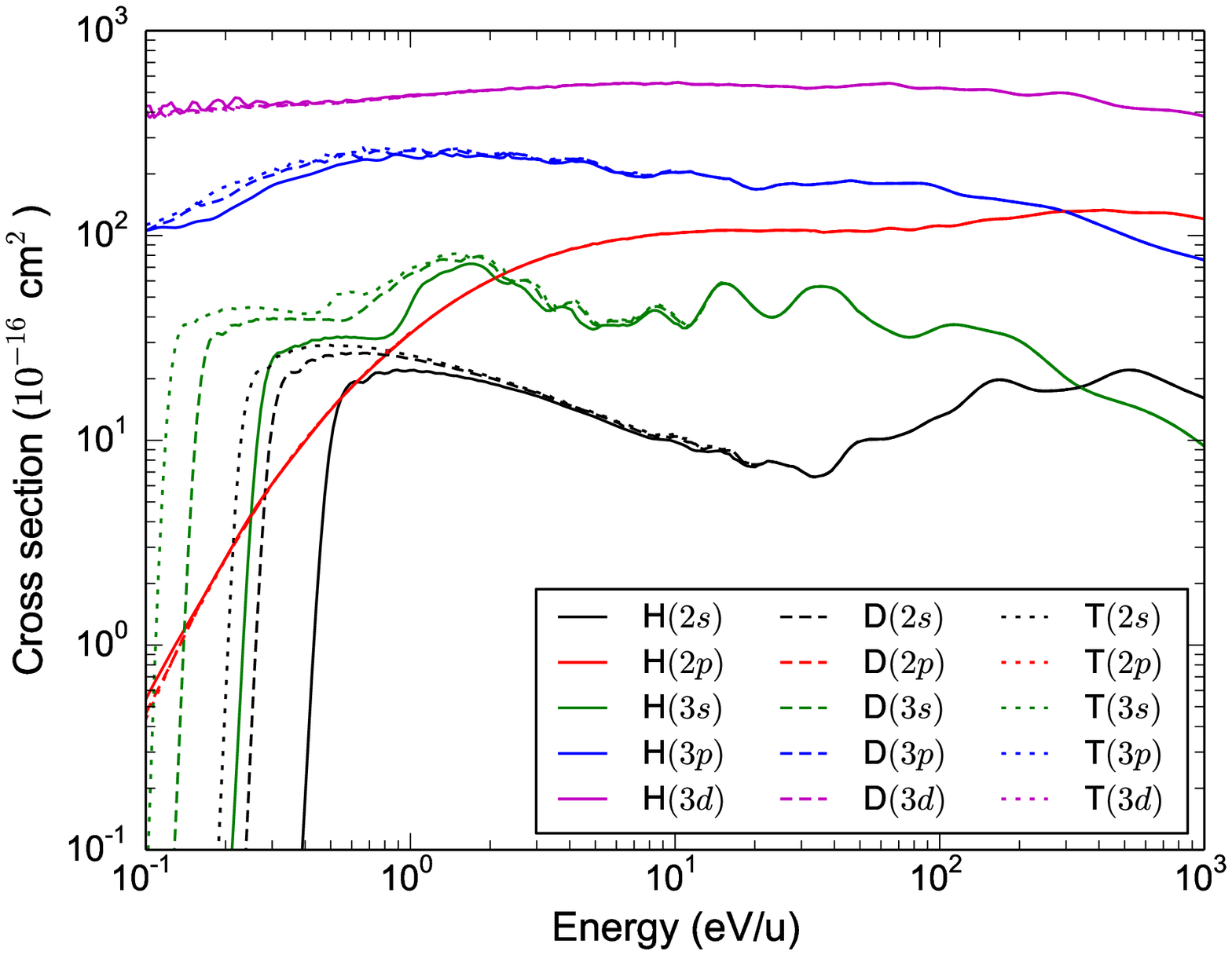}
\caption{Isotope effect on the total charge transfer cross sections for H/D/T($nl$) + He$^+$ collisions in singlet states.}
\label{fig_isotope_H_singlet_1}
\end{figure}

\begin{figure}
\includegraphics[,width=.5\textwidth]{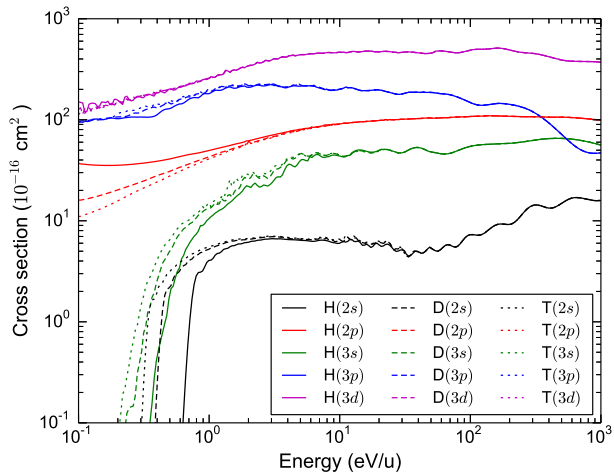}
\caption{Isotope effect on the total charge transfer cross sections for H/D/T($nl$) + He$^+$ collisions in triplet states.}
\label{fig_isotope_H_triplet_1}
\end{figure}

The charge transfer process (\ref{CX_1}) was previously investigated in Ref. \cite{Loreau2010b} for singlet states for collision energies between 0.25 eV/u and 150 eV/u. 
Several conclusions are worth recalling as they apply to the calculations presented here. 
First, it was observed that the dominant charge transfer cross sections occur when the initial and final states have the same principal quantum number, $n=n^\prime$. However, for higher values of $n$ than considered here manifolds with different $n$ are expected to interact more and more due to the increased density of states.
Second, the calculated cross sections are only weakly affected by the inclusion of higher-lying molecular states, thereby showing that the calculations are converged with respect to the number of states. 
Third, the cross section displayed a strong dependence in the principal and orbital quantum numbers $n$ and $l$ of the initial state, H($nl$) + He$^+(1s)$.

Fig. \ref{fig_comp_H_1} shows the cross sections for process (\ref{CX_1}) for the various initial states with $n=2,3$, which are obtained by summing the contributions of the different final states and molecular symmetries. 
The results for both singlet and triplet spin multiplicities are displayed as well as the total spin-averaged cross section, given by $\sigma_{\mathrm{tot}}=\frac{1}{4}\sigma(S=0)+\frac{3}{4}\sigma(S=1)$.

Up to 100 eV/u, we observe that the cross sections increase with the quantum numbers $n$ and $l$ of the initial state, both in singlet and in triplet states. This dependence in $n$ is expected as the density of states increases rapidly and the classical cross section for Rydberg states scales as $n^4$. However, at the highest energies investigated here this effect disappears to some extent: the cross section for H($3s$) + He$^+$ becomes smaller than the one for H($2s$) + He$^+$ in the singlet states, and the same is also true for H($3p$) and H($2p$) in both spin multiplicities. 

The cross sections for H($2s$) and H($3s)$ present a threshold corresponding to the energy required to reach the non-adiabatic couplings that control the charge transfer process from these states, as discussed in Ref. \cite{Loreau2011c}, both for singlet and triplet states. 
The cross sections from H($2p$) and H($3p)$ are larger than for the $s$ states over most of the energy range considered here. For these states, we also observe that above 20 eV/u, the cross sections for singlet and triplet states are similar. 
Finally, the cross section for H($3d$) dominates over the whole energy range. While at low energy the cross section for singlet states is much larger than for triplet states, they become comparable above 20 eV/u.

For charge transfer processes in which the electron capture leads to singlet and triplet states, it is well known that there can be large deviations to the statistical values \cite{Bliek1998,Stancil1997a,Wang2003}. Spin effects in H($1s$) + He$^+$ charge transfer collisions were previously shown to occur at low energy and persist until several tens of keV/u \cite{Hippler1989}.
Similarly, we also observe large differences between the cross sections in triplet and in singlet states. 
At the low energies considered here ($<1$ keV/u), the characteristics of the collision are dependent on the precise shape of the interaction potentials and the magnitude of the cross sections and the dynamics of charge transfer can be explained on the basis of the PECs as well as the location and shape of the non-adiabatic couplings \cite{Loreau2010b,Loreau2011c}. 
Since the PECs for the two spin multiplicities display distinct behaviors, as has been discussed elsewhere \cite{Loreau2010a,Loreau2011c}, differences between the cross sections for both spin multiplicities should be expected.
However, insights into the features that differentiate the two spin multiplicities can already be gained on the basis of the asymptotic energy defects (see also Table \ref{table_states}). 
For a given principal quantum number $n$, in the triplet manifold the states He($n\ ^3L$) + H$^+$ all lie below the states He$^+$ + H($nl$). 
The state He($n\ ^3S$) + H$^+$ is always the lowest state, followed by the other He($n\ ^3L$) states in order of increasing angular quantum number $L$. 
By contrast, in the singlet manifold, He($n\ ^1P$) + H$^+$ is always the highest state. In addition, while He($n\ ^1S$) + H$^+$ is still the lowest state, it is much closer in energy to the other electronic states with the same principal quantum number $n$.
Finally, it is worth noting that there are significant differences in the polarizabilities of the states He($n\ ^{1,3}L$): for instance, the polarizability of the $^1S$ states is larger than that of the $^3S$ states, while the polarizability of the $^1P$ states is larger than that of the $^3P$ states and of opposite sign. These properties of the atomic states will directly influence the shape of the PECs as well as the location of the avoided crossings that control the charge transfer dynamics.

In applications to modeling of laboratory plasmas, it is important to examine the influence of the hydrogen isotope effect on the charge transfer cross sections. The isotope effect occurs mainly at low collision energies, although it can remain present at energies up to tens of eV/u \cite{Stancil1995}, much larger than the isotopic mass shift. In most cases it was observed that the cross section decreases with an increase in isotope mass.
The effect of the isotopic substitution of H by D or T in process (\ref{CX_1}) is shown in Figure \ref{fig_isotope_H_singlet_1} for singlet states and in Figure \ref{fig_isotope_H_triplet_1} for triplet states. 
The largest isotope effect occurs for H($2s$) and H($3s$), both in the singlet and triplet manifolds. The large effect is a direct consequence of the threshold in the cross sections discussed above and the use of mass-scaled collision energy units. 
The origin and magnitude of the isotope effect from H($2p)$ has been discussed in Ref. \cite{Loreau2011c}. 
For H($3p$), we observe an increase of the cross section with the reduced mass of the system in the the singlet and triplet states. Interestingly, the isotope effect decreases at the lowest energy considered here, while the largest effects (about 40 \%) occurs at collision energies of 0.2 eV/u and 0.4 eV/u for singlet and triplet states, respectively.
In the case of H($3d$), there is a small (15\% at most) isotope effect below 1 eV/u in both singlet and triplet manifolds. 
In all cases, the isotope effect is limited to collision energies below 30 eV/u.

\subsection{Charge transfer in H$^+$ + He$(1snl\ ^{1,3}L)$ collisions}

We now focus on process (\ref{CX_2}), corresponding to the depopulation of an electronically-excited state of He in collisions with H$^+$ due to charge transfer. While the state-to-state cross sections for processes (\ref{CX_1}) and (\ref{CX_2}) are related through the detailed balance mechanism, it is not the case for the total cross sections. 

\begin{figure}
\includegraphics[,width=.5\textwidth]{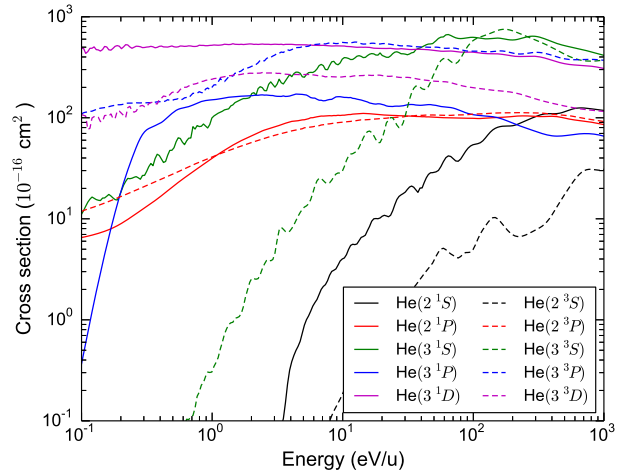}
\caption{Total charge transfer cross sections for H$^+$ + He$(1snl\ ^1L) \rightarrow$ H + He$^+$ collisions in singlet and triplet states.}
\label{fig_comp_He_1}
\end{figure}

Fig. \ref{fig_comp_He_1} displays the cross sections for process (\ref{CX_1}) for the five initial states with $n=2,3$ in the singlet and triplet manifolds. 
At the lowest energies the cross section increases with $n$ and $l$, as was observed for process (\ref{CX_1}). However, as higher energy this trend disappear completely. In particular, the cross section for He($3\ S)$ becomes dominant both in the singlet and triplet manifolds.

For the initial states He($1sns$) + H$^+$ and He($1snd$) + H$^+$, the cross section in singlet states is larger than that in triplet states. This is likely due to the larger asymptotic energy differences between the initial and final states of the collision in the triplet manifold of He($1snl$) states compared to the singlet manifold. 
This is particularly true for $S$ states, for which the triplet state is much lower in energy than the singlet state, leading to a smaller coupling with the other molecular states. 

We observe a different behavior for the initial states $^1P$ and $^3P$. 
For He($1s3p$) + H$^+$, the cross section for the singlet state is smaller than for the triplet state. As explained above, asymptotically the $3\ ^1P$ level lies above all other $n=3$ states, which is not the case of the $3\ ^3P$ state. Consequently, the corresponding molecular states interacts much less with the other $n=3$ states in the singlet than in the triplet manifolds, leading to smaller cross sections over the whole energy range considered here.
On the other hand, the cross sections for He($2 ^1P$) and He($2 ^3P$) seem to contradict this interpretation as they are similar in magnitude, and are almost equal above 20 eV/u. 
The explanation lies in the fact that for these initial states, charge transfer occurs differently in the $\Sigma^+$ and in the $\Pi$ molecular states. In the $\Sigma^+$ states, the cross section in the singlet manifold is much smaller than in the triplet manifold, for the reasons given above. In the $\Pi$ states, there are only two molecular states for $n=2$. Their PECs are almost parallel with a very wide non-adiabatic coupling, and the charge transfer process can be described using the Rosen-Zener-Demkov model \cite{Demkov1964,Bransden1992}. Since the asymptotic energy difference between He($1s2p$) + H$^+$ and He$^+(1s)$ + H($2p)$ is much larger in the triplet states, the charge transfer cross section will be smaller. The combination of these two effects results in similar cross sections for singlet and triplet states. 

\begin{figure}
\includegraphics[width=.5\textwidth]{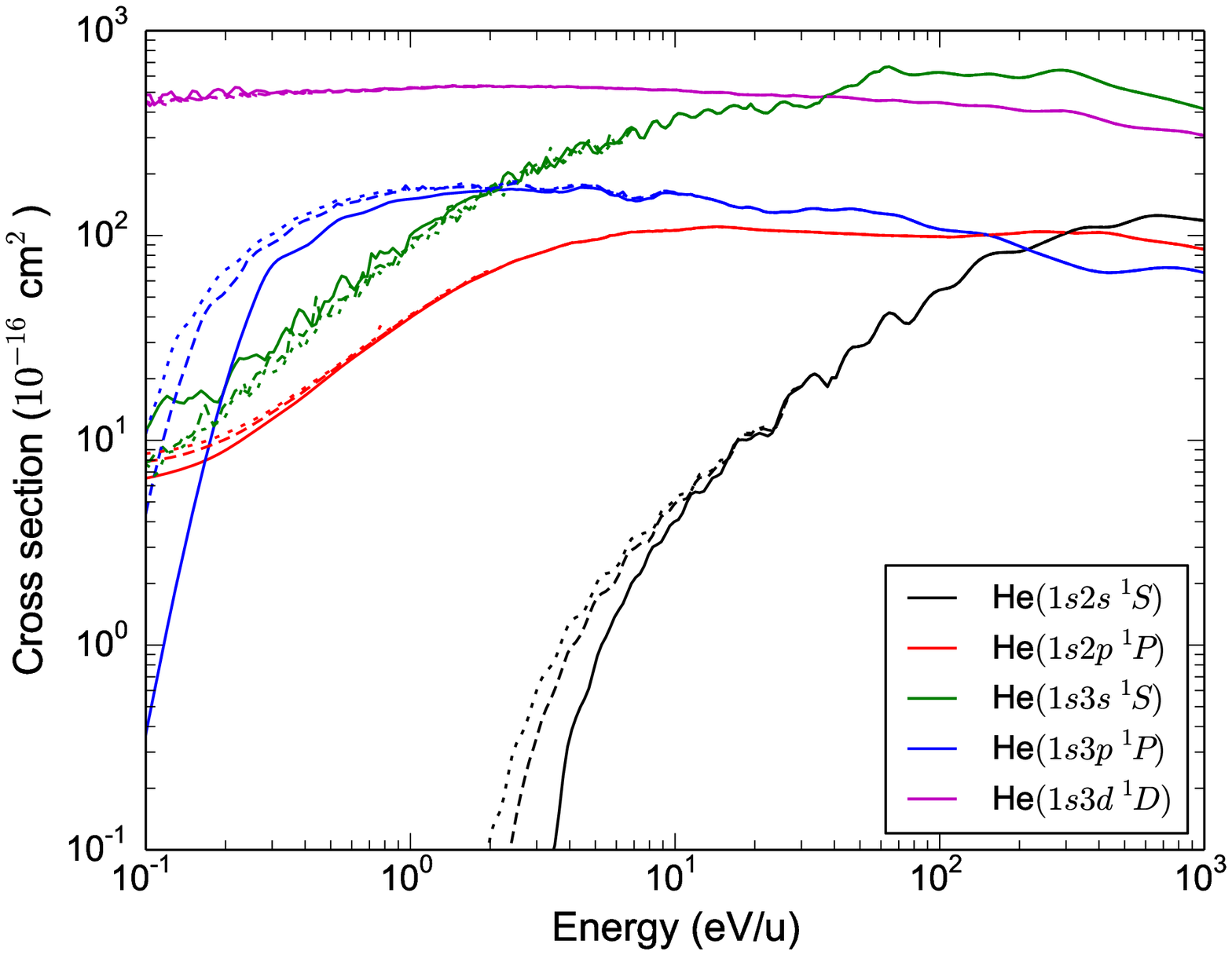}
\caption{Charge transfer cross sections for H$^+$/D$^+$/T$^+$ + He$(1snl\ ^1L)$ collisions in singlet states. Full lines: collisions with H$^+$; dashed lines: collisions with D$^+$; dotted lines: collisions with T$^+$.}

\label{fig_isotope_He_singlet_1}
\end{figure}

\begin{figure}
\includegraphics[width=.5\textwidth]{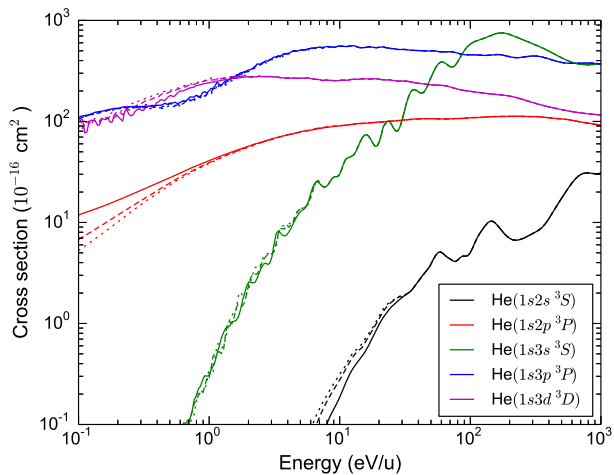}
\caption{Charge transfer cross sections for H$^+$/D$^+$/T$^+$ + He$(1snl\ ^1L)$ collisions in triplet states. Full lines: collisions with H$^+$; dashed lines: collisions with D$^+$; dotted lines: collisions with T$^+$.}
\label{fig_isotope_He_triplet_1}
\end{figure}

The isotope effect where H$^+$ is substituted by D$^+$ or T$^+$ in reaction (\ref{CX_2}) is illustrated in Fig. \ref{fig_isotope_He_singlet_1} for singlet states and in Fig. \ref{fig_isotope_He_triplet_1} for triplet states. 
In the singlet states, we observe a very small effect on the He($2\ ^1P$) and He($3\ ^1D$) cross sections limited to collision energies below 1 eV/u. For the states He($2\ ^1S$) and He($3\ ^1P$), at a given energy (in mass-scaled units) the isotope effect results in an increase of the cross section with increasing mass of the isotope. For He($3\ ^1S$), the cross section decreases with increasing isotope mass. 
In all cases, the isotope effect vanishes at collision energies above 20 eV/u. 

In the triplet states, there is no substantial difference between the three isotopes for the states He($3\ ^3S$), He($3\ ^3P$), and He($3\ ^3D$) with variations of 40\% at most. For He($2\ ^3P$), we observe a decrease of the cross section with increasing isotope mass, with deviations that reach a factor of 2 at 0.1 eV/u. The largest isotope effect occurs for He($2\ ^3S$), for which the cross section increases by orders of magnitude with increasing isotope mass. However, it should be noted that at the energies at which the isotope effect is appreciable (below 5 eV/u) the charge transfer cross section for this state is extremely small.

The process (\ref{CX_2}) has been previously studied theoretically \cite{Chibisov2001,Chibisov2002} in the energy range between 2 eV/u and 5 keV/u. 
The authors used an atomic orbital semi-classical approach with a linear trajectory for the nuclei that takes into account the coupling between the molecular states due to the Stark effect at large internuclear distances ($R\geq 20$ a.u.). There are large discrepancies between the cross sections presented in Refs. \cite{Chibisov2001,Chibisov2002} and the results discussed above, both in the magnitude of the cross sections but also in their dependence on the collision energy. Some of these discrepancies were already discussed in Ref. \cite{Loreau2010b}. In general the charge transfer cross sections presented in Refs. \cite{Chibisov2001,Chibisov2002} seem to be strongly overestimated, as was also noted by Liu {\it et al.} \cite{Liu2012b} at energies above 2 keV/u.
In addition, we note that the cross sections for He($3\ ^{1,3}S$) are predicted to increase with decreasing energy, in complete contradiction with the results presented in Fig. \ref{fig_comp_He_1} and our analysis. Moreover, the cross sections for both states have similar magnitudes below 40 eV/u, which does not appear in our calculations. 
For the He($3\ ^{1,3}P$) and He($3\ ^{1,3}D$) states, the agreement is somewhat better, although there are large discrepancies regarding the magnitude of the cross sections. 
This shows the limitation of the semi-classical method of Ref. \cite{Chibisov2001,Chibisov2002}.

\section{Conclusions and perspectives}

We have computed the total charge transfer cross sections for collisions of H($2s$), H($2p$), H($3s$), H($3p$), and H($3d$) with He$(1s)$, as well as for collisions of He($2\ ^{1,3}S$), He($2\ ^{1,3}P$), He($3\ ^{1,3}S$), He($3\ ^{1,3}P$), and He($3\ ^{1,3}D$) with H$^+$. The cross sections were calculated for collision energies form 0.1 eV/u up to 1 keV/u using both a quantum-mechanical method at low collision energies and a semi-classical approach at higher energies. Both methods are based on a molecular approach.

As expected, we observed an increase of the cross section with the quantum number $n$. For collisions of H($nl$) with He$^+$, there is an additional dependence in $l$, the cross section increasing with increasing $l$. For He($n\ ^{1,3}L$) + H$^+$ collisions, this dependence does not appear except at the lowest energies considered here. 
The relative magnitude of the cross sections in singlet and triplets manifolds was interpreted on the basis of the asymptotic energy defects, although at very low energy the shape of the potential energy curves as well as the strength and location of the non-adiabatic couplings are obviously critical factors. 

The effect of isotopic substitution on charge transfer cross sections was investigated. The isotope effect occurs at energies below to 30 eV/u and generally increases with decreasing energies. The cross sections that present thresholds are the most affected by isotopic substitution, the effect reaching orders of magnitude. In the other cases, the effect is much smaller although it can reach a factor of 4 in the case of H$(2p)$ + He$^+$.

In the future, the cross sections presented in this work will be used to improve modelling of laboratory plasmas such as the collisional-radiative models that are used to study He lines in fusion plasmas.
Forward atomic modeling of helium gas-puff emission is a powerful tool for assessing and estimating line intensities used for line-ratio spectroscopy diagnostics in laboratory plasmas \cite{Bertschinger2005,Munoz2012}. 1-D kinetic and time-dependent models have already been developed and employed on helium simulations and diagnostic applications \cite{Munoz2012,Munoz2016}. Although consistent results have been obtained using the standard 667.9, 706.7, and 728.3 nm He I lines, some discrepancies have been found when using other He I lines in electron temperature and density diagnostics on the ASDEX-Upgrade experiment \cite{Munoz2016}.

The diagnostic methodology consists of puffing helium gas into the plasma, where the populations of the atoms in the ground and excited states are affected by various atomic processes. These atomic interactions are a function of electron and ion temperatures and densities. By modeling the different populating processes, it is possible to predict the emission from different lines that are sensitive to local plasma conditions in order to use them for diagnostics. Due to the singlet and triplet (metastable) spin systems found in He I, relaxation times of different excited levels can become significant for lower when low electron density conditions exist.  A hybrid time dependent/independent collisional radiative model that takes into account the high Rydberg contributions and transient effects on the helium populations has already been developed \cite{Munoz2016}; however, it does not include the contributions of charge exchange or electron recombination processes in excited states.
In order to extend the applications of this powerful diagnostic technique to various plasma regimes by using different He I emission lines, these atomic processes must be included in the model. Charge-exchange between proton-helium collisions have already been assessed by simulations in the SOL-Edge regions of NSTX; however, charge-exchange between H and He$^+$ collisions may become significant for different plasma conditions, particularly for the H($2s$) and H($2p$), as well as H($3s$), H($3p$), and H($3d$) states \cite{Rosmej2006,Loreau2010b}. These charge-exchange processes can significantly affect the populations of the singlet and triplet (metastable) spin systems of He I, which in turn will have a strong effects on other He I emission lines used for diagnostic purposes \cite{Munoz2012}.

An important application of the set of calculated cross sections presented in this work consists in the development of a 1-D kinetic collisional and radiative model that will include a more comprehensive set of atomic collision processes, such as: electron-impact excitation and ionization, electron recombination, charge transfer between H$^+$/He and H/He$^+$, as well as high Rydberg contributions to the lower populations. This model will be designed to individually assess the contributions of each of the atomic processes to line emission, enabling the design of a new generation of helium line-ratio diagnostics under different plasma conditions.

\acknowledgments

J.L. thanks V. Kharchenko and M. Desouter-Lecomte for useful discussions.
This work was supported by the I.I.S.N. instrument (grant 4.4504.10) of the Fonds de la Recherche Scientifique - FNRS.
Computational resources have been provided by the Shared ICT Services Centre of the Universit\'e libre de Bruxelles.

\end{document}